# Structure of Liquid Nitromethane: Comparison of Simulation and Diffraction Studies

Tünde Megyes[a)], Szabolcs Bálint, Tamás Grósz, Tamás Radnai, Imre Bakó

*Institute of Structural Chemistry, Chemical Research Center of the Hungarian Academy of Sciences, Pusztaszeri út 59-67, H-1025 Budapest, Hungary*

László Almásy

*Research Institute for Solid State Physics and Optics, P.O. Box 49, Budapest 1525, Hungary*

**Abstract**

Simulation (MD and Car-Parrinello) and diffraction (X-ray and neutron) studies are compared on nitromethane with aiming at the determination of the liquid structure. Beyond that, the capabilities of the methods to describe liquid structure are discussed. For the studied liquid, the diffraction methods are performing very well in determination of intramolecular structure, but they do not give detailed structural information on the intermolecular structure. The good agreement between the diffraction experiments and the results of molecular dynamics simulations justify the use of simulations for the more detailed description of the liquid structure using partial radial distribution functions and orientational correlation functions. Liquid nitromethane is described as a molecular liquid without strong intermolecular interactions like hydrogen bonding, but with detectable orientational correlations resulting in preferential anti-parallel order of the neighboring molecules.

---

[a)] Author to whom correspondence should be addressed. Electronic mail: megyes@chemres.hu

## 1. Introduction

Nitromethane is one of the simplest nitrogen-containing molecules belonging to the nitro compounds that are of great interest to the high explosives community. It has been heavily studied due to its use as a fuel and as a prototype molecule for a class of high-energy materials and perhaps plays a role in the atmosphere[1] as well. Shock wave induced decomposition of nitromethane has been subject to many investigations.[2] Detailed studies of formation of negative ions,[3] electron transfer processes[4] and gas phase solvation processes[5] in nitromethane were also reported.

Nitromethane is a small highly polar molecule (dipole moment 3.46 D) for which pair association in the liquid state has been claimed several times in the literature.[6] Recently Cataliotti et al. concluded from their IR and Raman spectroscopic experiments that liquid nitromethane has molecules in monomeric state and is not associated in pairs as reported earlier.[7]

There exist various theoretical studies on the structure and properties of nitromethane. Byrd et al.[8] performed *ab initio* study of solid nitromethane on four energetic molecular crystals; they have found that the lattice vectors determined display large errors possibly due to the lack of van der Waals forces in functionals applied in current DFT theories and further development of the methods of DFT was suggested. Structural and vibrational properties of solid nitromethane were also studied by the DFT method.[9] Correlated calculation of the interaction in the nitromethane dimer was performed on a fixed monomer configuration.[10] *Ab initio* study of nitromethane dimer and trimer[11] was performed by employing the density functional theory B3LYP method. For the optimized structure of nitromethane dimer the strength of C–H···O–N bond ranges from −9.0 to −12.4 kJ·mol$^{−1}$ at MP2 level while the B3LYP method underestimates the interaction strength compared with MP2 method.

Molecular dynamics simulations were performed on both crystalline and glassy nitromethane,[12] on melting of nitromethane,[13] on nitromethane nanoparticles[14] and on liquid nitromethane.[15,16,17]

Diffraction techniques serve, in principle, as direct method to determine the structure of the matter in condensed state. The crystalline structure of nitromethane has been determined using single crystal X-ray and neutron diffraction.[18] To the best of our knowledge, diffraction study of liquid nitromethane was not performed yet. In the present study we conducted X-ray and neutron diffraction experiments for the first time on nitromethane, in combination with molecular dynamics simulation that uses the intermolecular interaction potential of Sorescu et al.[16] and a Car-Parrinello[19] simulation. To learn more about the mutual orientation of nitromethane molecules we have also performed an *ab initio* study of nitromethane dimer on MP2 level.

The paper is organized as follows: In the 'methods' section we describe the quantum chemical calculations, and the simulation details and after the X-ray and neutron measurements and data treatments. In 'results and discussion' we present the experimental and theoretical results and compare them before 'conclusions' summarizes the main results.

## 2. Computational Details

### 2.1. Quantum chemical calculations

All calculations were performed by using the Gaussian 03 program suite[20] at DFT/B3LYP and MP2 level of theory using a 6-311+G** basis set. The behavior of the calculated stationary points was characterized by their harmonic vibrational frequencies. The interaction energies for each minimum were corrected for basis set superposition error (BSSE) with the full counterpoise (CP) procedure, resulting in a more reliable estimate of the interaction energy.[21] The magnitude of the BSSE correction at the energy minimum configuration is about 7-8% and 25-30% of the total interaction energy at B3LYP and MP2 level of theory, respectively. In principle, since the BSSE causes the intermolecular interactions to be too attractive, the CP correction is expected to make the complexes less stable.[22]

*2.2. Molecular Dynamics Simulation*

We have performed a classical MD simulation in the NVT ensemble. The simulation box contained 500 rigid nitromethane molecules. The intermolecular interactions were defined as by Buckingham potentials, developed by Sorescu et al.[16] The side length of the cube was 35.56 Å. During the 25,000 time steps of equilibration the Nosé–Hoover thermostat was used to control the temperature in the DLPOLY 2.15 software.[23] The simulation was performed for 200,000 time steps leading to the total time of 400 ps.

*2.3. Car-Parrinello Simulation*

An other simulation was performed by using the Car-Parrinello[24] *ab initio* molecular dynamics scheme. The valence electronic wave functions were expanded in plane waves with a 25 Ry cutoff and the valence-core interaction was described by the Vanderbilt ultrasoft pseudopotentials.

A liquid-like system was made up of 32 $CD_3NO_2$ molecules enclosed in a cubic box with a density matching that of the experiment. A classical molecular dynamics simulation using the Sorescu et al.[16] potential model for liquid nitromethane was used to generate the initial configuration. The CPMD simulation was run using a time step of $\Delta t=0.163$ fs and the fictious electron mass ($\mu$) was 600 au. A continuous trajectory of 15 ps was obtained in the microcanonical ensemble with the last 6 ps, and it was used for the computation of average properties. The temperature was set to 300 K.

Recently a series of Car-Parrinello and Bohr-Oppenheimer molecular dynamic simulations were carried out for liquid water to investigate the reproducibility of this method.[25,26,27] They showed that for structural properties the size effects are rather small, but care is required in the choice of an appropriate electron mass. In Car-Parrinello simulations it is important to maintain an adiabatic separation between the electronic and ionic degrees of freedom. In our case this could be achieved using a mass ratio $\mu/M = 1/6$ (M: smallest atomic mass of the system). We did not find any energy drift in our simulation, and the conserved energy fluctuation was about $10^{-7}$% (~$10^{-4}$ au). In order to verify the accuracy of our model we analyzed the structure of the nitromethane molecule and the nitromethane dimer at different levels of theory. Our calculations, using plane-wave basis set produced results, which agree very well with the high level *ab initio* quantum chemical calculation findings.

The self-diffusion coefficient is a sensitive indicator of the accuracy of the applied model. The self-diffusion coefficient was estimated from the Einstein relation, which uses the long time limit of the mean square displacements of the center of mass of the molecule. The value predicted from the simulation is about $0.9 \cdot 10^{-9}$ m$^2$/s, which is about 70% smaller than the classical simulation value for pure $CH_3NO_2$ liquid at 298 K ($1.52 \cdot 10^{-9}$ m$^2$/s). Such a difference may be an artifact caused by the small size of the simulation box, as it was already proven in earlier studies. Thus, applying the same classical potential model for a system containing 32 molecules, the calculated diffusion constant was about $1.1 \cdot 10^{-9}$ m$^2$/s. Both these values are still lower than the experimental result, which is 2.3 and 2.4 $10^{-9}$ m$^2$/s for proteated and deuterated nitromethane, respectively, at 298 K.[28] Such deviations are not rare in simulations of liquids when the potentials are not fully optimized for reproducing thermodynamic properties.

**3. Details of the experimental studies**

*3.1. X-ray diffraction measurement and method of structural analysis*

X-ray diffraction measurement was carried out on liquid nitromethane, anhydrous, special grade, produced by Aldrich. The physical properties of the nitromethane were: density $\rho$ = 1.14 g·cm$^{-3}$, linear X-ray absorption coefficient $\mu$ = 1.0329 cm$^{-1}$, atomic number density $\rho_0$ = 0.0787 $10^{-24}$ cm$^3$.

The X-ray scattering measurements were performed at room temperature (24±1°C), with a Philips X'Pert goniometer in a vertical Bragg-Brentano geometry with a pyrographite monochromator in the scattered beam and proportional detector using MoK$\alpha$ radiation ($\lambda$=0.7107 Å). Quartz capillaries (1.5 mm diameter, 0.01 mm wall thickness) were used as the liquid sample holder. The scattering angle range of measurement spanned over 1.28 ≤2Θ≤130.2° corresponding to a range of 0.2Å$^{-1}$≤$k$≤16.06 Å$^{-1}$ of the scattering variable $k$= (4$\pi$/$\lambda$)·sinΘ. Over 100,000 counts were collected at each angle in $\Delta$k≈0.05 Å$^{-1}$ steps.

Background and absorption corrections were applied based on an algorithm reported by Paalman and Pings[29] for cylindrical sample holders. This algorithm assumes that significant coupling does not occur between the sample and cell,[30] therefore the experimentally observed intensities are considered as linear combination of an independent component from the confined sample and a component from the sample cell. The correction procedure was applied using in house software written in a Fortran language. The polarization and Compton scattering corrections were applied using standard methods given in earlier works.[31]

The experimental structure function is defined as:

$$h(k) = I(k) - \sum_{\alpha} x_{\alpha} f_{\alpha}^2(k) M(k) \qquad (1)$$

where $I(k)$ is the corrected coherent intensity of the scattered beam normalized to electron units[32] $f_{\alpha}(k)$ and $x_{\alpha}$ are the scattering amplitude and mole fraction for a type of $\alpha$ particle, respectively; $M(k)$ is the modification function, $1/[\Sigma x_{\alpha} f_{\alpha}(k)]^2$. The coherent scattering amplitudes were calculated as previously described.[31] The nitromethane molecules were treated in atomic representation, and the necessary parameters were taken from the International Tables for X-ray Crystallography.[33]

The experimental radial distribution function was computed from the structure function $h(k)$ by Fourier transformation according to equation 2

$$g(r) = 1 + \frac{1}{2\pi^2 r \rho_0} \int_{k_{min}}^{k_{max}} k h(k) \sin(kr) \, dk \qquad (2)$$

where $r$ is the interatomic distance, $k_{min}$ and $k_{max}$ are the lower and upper limits of the experimental data, $\rho_0$ is the atomic number density. After repeated Fourier transformations the non-physical peaks present in the $g(r)$ at small $r$ values were removed, and the structure function was corrected for residual systematic errors.[34]

In order to characterize the structure of the liquid, as a first step, a visual evaluation and a preliminary semi-quantitative analysis of the observed structure functions $kh(k)$ and radial distribution functions $g(r)$ were performed. Further on, the observed data were analyzed by geometrical model constructions and fitting the model structure functions to the corresponding experimental ones by the non-linear least-squares method. The fitting strategy was previously described in refs. 31.

*3.2. Neutron diffraction measurements*

Neutron diffraction experiment on liquid nitromethane $CD_3NO_2$ (99%) was carried out on the 7C2 diffractometer of the Laboratoire Léon Brillouin CEA-SACLAY in a range of $0.3 \leq k \leq 15.3$ Å$^{-1}$. The liquid was kept in a vanadium container of 6 mm in diameter and 0.1-mm wall thickness. The incident neutron wavelength was 0.70 Å. For standard corrections and normalization procedures, additional runs (vanadium bar, cadmium bar, empty container and background) were also performed. The raw diffraction data were corrected for background, container and sample absorption, multiple scattering, and then the intensities were normalized to absolute scale, by using scattering of a vanadium sample. A more detailed description of the correction procedure can be found elsewhere.[35]

The conversion of the observed total cross section $d\sigma/d\omega$ to an *r*-space representation was performed with the MCGR method[36] (*'Monte Carlo treatment of the experimental radial distribution function'*). In this method, the radial distribution functions, either total or partial, are generated numerically and modified by a stepwise random Monte Carlo process until its inverse Fourier transform agrees with the experimentally scattering cross section within the limits of experimental error.

## 4. Results and discussion

*4.1 Quantum chemical calculations*

The parameters obtained for the optimized structure of nitromethane molecule with the methods employed in this work together with their experimental counterparts[37,38] are presented in Table 1. It can be observed that all geometrical parameters obtained from two different levels of theory are in good agreement with experimental results. The two different conformers *eclipsed* and *staggered* of nitromethane molecule are shown in Figure 1. These conformers have nearly the same total energy. In the first minima of the ONCH$_1$ dihedral angle is about 90º, but in the other one it is about 180º. The rotation barrier between these two minima is less then 0.01 kcal/mol, so it can be concluded that in gas phase the rotation of NO$_2$ group around the C–N bond is almost free. These results agree very well with the earlier findings.[39,40] The rotational barrier between this two conformers, at T = 0 K, in solid state is about 0.54 kcal/mol[41] and in gas phase is about 0.006 kcal/mol.[38,42]

The optimized geometry for the dimer of nitromethane molecules is shifted anti-parallel as it is shown in Figure 2. Table 2 reports resulting parameter values for the anti-parallel structure. Comparison between the geometries of the isolated monomer and the monomer geometry in the dimer shows no significant change. Table 2 reveals that the calculated O···H distance is about 2.4-2.5 Å, which is slightly less than the sum of the van der Waals radii of hydrogen and oxygen atoms. From this distance it may be suggested that the nitromethane molecules interact with each other in the dimer by weak C-H···O interaction. In previous works[10,11] was found that in the nitromethane dimer weak C-H···O interaction held together the two molecules. In our case the BSSE corrected interaction energy is about 4.16 and 4.45 kcal/mol at B3LYP and MP2 level, respectively. It has already been shown, that the weak H-bonded interaction, if appears, can be detected by Bader analysis (atom in molecule method, AIM), IR spectroscopy (change of CH stretching vibration frequency) and natural orbital analysis (NBO).

Eight AIM criteria have been proposed before to study and characterize hydrogen bonds and decide, whether they are conventional or CH···O bond.[43,44] These properties are connected to the topology of the electron density. When applying these criteria in our case, however, we cannot detect any bond critical point along the CH···O line.

The characteristic stretching frequencies of methyl group in monomer and dimer are shown in Table 1 and 2. There are no significant differences between the $CH_3$ stretching frequencies in monomer and dimer structure.

It has also been shown that the NBO analysis is a useful tool to characterize the electron transfer processes from the proton acceptor to the proton donor.[45] Reed et al.[46] investigated several typical H-bonded systems, demonstrating charge transfer from the lone pairs of the proton acceptors to the antibonding orbitals of proton donor. Especially helpful for the characterization of the hydrogen bond is the second order perturbation energy lowering due to the interaction of the donor and acceptor orbitals. In our case this value of energy lowering is very small (less than 0.1 kcal/mol, but for a typical H-bonded dimer is about 1-2 kcal/mol). The conclusion from the above-mentioned considerations is that in the nitromethane dimer the CH···O interaction cannot be considered as H-bonded interaction.

*4.2. MD and Car-Parrinello Simulation Results*

*4.2.1. Radial Pair Distribution Functions (RDF's)*

The structure of liquid nitromethane was analyzed in terms of radial distribution functions (RDF's), denoted as $g_{\alpha\beta}(r)$, for the various atom-atom pairs. The corresponding running integration numbers $n_{\alpha\beta}(r)$ are defined by:

$$n_{\alpha\beta}(r) = 4\pi\rho_\beta \int_0^r g_{\alpha\beta}(r) r^2 dr \qquad (3)$$

The value of this integral up to the first minimum ($r_{m1}$) in $g(r)$ is the number of coordinating atoms of type $\beta$ around atoms of type $\alpha$ and $\rho_\beta$ is the number density of the atoms of type $\beta$ at a distance $r$. The molecular dynamics simulation produces individual pair distribution functions for each of the interactions and these can be analyzed to get an insight into the arrangement of the molecules in the liquid. The RDFs of the liquid nitromethane obtained from classical and Car-Parrinello simulations are presented in Figure 3, and the characteristic values of RDFs obtained by classical method are given in Table 3.

The essential features of the RDFs in the two types of simulations agree with each other. A small systematic shift of all peaks towards larger $r$ is apparent for the CP results. As the statistical accuracy of the classical MD simulation is superior to the Car-Parrinello simulation, we will use further the values obtained from the classical MD. The presently obtained RDFs can also be compared to the results of earlier classical computer simulations. We found our RDF functions to be rather similar to those obtained in simulations by Alper et al.[15] and Sorescu et al.[16]

The analysis of RDFs containing the intramolecular interactions obtained by Car-Parrinello method revealed that the molecular structure in the liquid state does not change significantly compared to the gas phase. It is well known that in the gas phase the rotation of the $CH_3$ group is nearly free and in the solid state there is a barrier of 0.6-0.86 kcal/mol between the two conformers of the nitromethane molecule.[39]

The angle distribution of the HCNO dihedral angle (Figure 4), as obtained from the Car-Parrinello simulation, shows a uniform distribution between 0 and $\pi/6$, indicating that in the liquid state the rotation of $CH_3$ group is free.

In the crystalline state[18a] of nitromethane there are about 18 short CH···O distances per molecule ($r_{CH···O} < 2.6$ Å, which may correspond to the intermolecular CH···O bonds). In the O···H partial radial distribution function a first small peak at around 2.92 Å and an additional one at 4.23 Å can be found (Figure 3). The featureless $g_{OH}(r)$ function in the range of 2-2.6 Å serves as a strong evidence for the non existence of a CH···O type hydrogen bond in the liquid state. Seminario et al.[17] performed density functional/molecular dynamics study of liquid nitromethane and their results clearly indicate the presence of CH···O type hydrogen bond. The reason for this discrepancy could be in the difference in the model potentials. We remark that the present Car Parrinello simulation does not use any empirical potential parameter and gives the first peak of $g_{OH}(r)$ at 3.1 Å. This distance is even longer than that obtained by molecular dynamics simulation, and does not support the existence of CH···O type hydrogen bond in liquid.

*4.2.2. Orientation of the nitromethane molecules*

The orientations of the neighboring molecules can be characterized by the angle dependent radial distribution function. Specifically, we have calculated the angle between the dipole moment vector of center and neighboring molecules as a function of N···N distance. The calculated angular radial distribution functions are shown in Figure 5. It can be seen that in liquid nitromethane only the first nearest neighbors tend to be oriented in an anti-parallel form. For N···N distances longer than 4.0 Å, the angular distribution is almost constant, indicating that the preference in orientation is lost very quickly.

The correlation of the relative orientation of the molecules in the liquid state can be characterized by the coefficients of a spherical harmonic expansion of the orientational radial distribution function. The details of the spherical harmonics expansion as well as the orientational correlations function's calculation using this expansion are given elsewhere.[47] Here we summarize the technique, following the notation used by Grey and Gubbins.[48] According to this formalism the total correlation function g(r, $\omega_1$, $\omega_2$) can be expanded by the following equation:

$$g(r,\omega_1,\omega_2) = \sum_{l_1,l_2,l} \sum_{n_1,n_2} g(l_1 l_2 l : n_1 n_2 : r)\Phi_{l_1 l_2 l, n_1 n_2}(\omega_1,\omega_2) \tag{4}$$

where $\Phi_{l_1 l_2 l, n_1 n_2}(\omega_1 \omega_2)$ are the generalized spherical harmonics functions, $g(l_1 l_2 l : n_1 n_2 : r)$ functions are the $r$ dependent expansion coefficients, which can be written into the following form:

$$g(l_1 l_2 l : n_1 n_2 : r) = 4\pi \frac{(2l_1+1)(2l_2+1)}{(2l+1)} g_{cc}(r)\Phi^*_{l_1 l_2 l, n_1 n_2}(\omega_1 \omega_2) \tag{5}$$

where $g_{cc}(r)$ is the centre-centre (in our case the N···N) radial distribution function and $\Phi^*_{l_1 l_2 l, n_1 n_2}(\omega_1 \omega_2)$ are the complex conjugate of the generalized spherical harmonics functions.

It is instructive to compare the orientational correlations in nitromethane and another similar liquid, acetonitrile ($CH_3CN$). Both liquids have similar dipole moments (acetonitrile 3.44, nitromethane 3.46 Debye) and possess an identical hydrophobic group (CH$_3$). Simulations and experiments indicated, that in the liquid acetonitrile the neighboring molecules have a strong preference for anti-parallel and slight preference for parallel head-to-tail configurations.[49]

The orientational coefficients for the total correlation function were determined in liquid nitromethane from the present molecular dynamics simulation. Some of these coefficients are summarized in Figure 6 together with the coefficients for liquid acetonitrile calculated from a classical simulation using six sites Bohm[50] model.

The $g_{00}^{110}$ coefficient is proportional to $<-\cos(\Theta)>$, where $\Theta$ is the angle between the dipolar axes of the two molecules. This function shows a maximum around 4.0 Å, which is a sign of the preference of the molecular dipoles in an anti-parallel alignment. This term is very similar to all of the investigated cases (classical MD of nitromethane and acetonitrile, and CPMD of nitromethane, not shown in Figure 5). The other term, which is connected to the angle of the dipole vectors, only is the $g_{00}^{220}$. In its case there is no significant difference between the coefficients of liquid nitromethane and acetonitrile. The $g_{00}^{101}$ and $g_{00}^{202}$ terms are proportional with the $<\cos(\Phi)>$ and $<\cos^2(\Phi)>$ terms, respectively, where $\Phi$ is the angle between the dipole vector of a central molecule and the vector connecting the two molecules. It is worth noting that we have calculated the orientational correlation functions for Car Parrinello simulation as well (not shown here), and they agree well with those obtained from classical simulation of nitromethane.

The significant difference in the $g_{00}^{101}$ and $g_{00}^{202}$ terms between the liquid nitromethane and acetonitrile can be explained with the help of angular radial distribution function, when the angle distribution (cos($\Phi$), angle between the dipole moment of central molecule and the center-center vector) as a function of N⋯N distance was calculated (Figure 7). Figure 7a shows that in liquid acetonitrile the cos(Φ) angle distribution has a symmetrical shape around 0, corresponding to the energy minimum configuration of acetonitrile dimer. In the case of liquid nitromethane, Figure 7b, the center of cos(Φ) distribution is around 0.25, corresponding to the shifted anti-parallel orientation of nitromethane dimer.

On the basis of simulation study it can be concluded that, the local structure of liquid nitromethane is determined by dipolar forces, with a slight preference to anti-parallel and no preference to parallel or head-to-tail configurations. It should be mentioned, that we have obtained this structural information based on the detailed analysis of the molecular dynamics simulations. The previous MD studies,[15,16] did not provide any comprehensive structural information, for they were focused mainly on testing the newly developed model potentials and comparing their predictions with the experimentally accessible vibrational quantities and the overall thermodynamic behavior of the liquid. Given the obvious limitation of the information contained in angle-averaged RDFs, it is evident from the present study that for deeper understanding of the liquid structure by using computer simulations, it is necessary to analyze the obtained configurations as deeply as possible.

*4.3. Structural results from X-ray diffraction*

At first a semi-quantitative analysis was done at the level of the radial distribution functions; as a second step a least square fitting method was used to determine the intra- and intermolecular structural parameters. The average scattering weighting factors of the different partial distribution functions were: C–C: 0.04, C–N: 0.08, C–O: 0.18, N–O: 0.21, N–N: 0.05 and O–O: 0.25. After examination of the weights of the contributions to the structure function one contribution for each type of interatomic distance listed in Table 2 was involved in the fitting procedure.

The experimental and theoretical X-ray structure functions, derived from experiment are shown in Figure 8, and the radial distribution functions are shown in Figure 9.

For the first peak centered around 1.20 Å, intramolecular C–N and N–O interactions are responsible. The small second peak can be observed in the range 1.85-2.70 Å. This peak can be assigned to C–O and O⋯O distances. Another broad peak appears in the range 3.505.85 Å. This peak is difficult to resolve because of its complexity, and can be attributed to various intermolecular atom-pair interactions.

The structural parameters obtained from the least-squares fit of the structure functions $kh(k)$ shown in Figure 8 are given in Table 4. The fitting procedure resulted in 1.49±0.01 Å and 1.22±0.01 Å for the intramolecular C–N and the N–O distances, respectively.The O⋯O distance was found to be 2.17±0.02 Å and C–O distances 2.32±0.01 Å. Once the intramolecular structure was found, various models were tested to determine the intermolecular structure of the liquid. Trial models containing nitromethane dimers in parallel and/or anti-parallel orientations failed completely. Then a decision has been made not to assume any initial intermolecular geometrical picture for the structure at the beginning of the fitting procedure. Due to their low contribution to the total scattering picture, intermolecular interactions (C⋯C, C⋯N, N⋯N) could not have been determined. The three other intermolecular distances, C⋯O, N⋯O and O⋯O, contribute with nearly similar weights to the total scattering intensity and it is not possible to resolve the interactions one by one. Consequently, no suggestions could have been made on the intermolecular structure of liquid nitromethane on the basis of X-ray diffraction alone. The solution for this problem might be to perform a comparison of the intermolecular radial distribution functions obtained by X-ray diffraction and molecular dynamics simulation. (see section 5.)

*4.4. Structural results from neutron diffraction*

The total cross section, dσ/dω, of liquid deuterated nitromethane and the structure function, *I(Q)*, together with the results of MCGR procedure are presented in Figure 10. The total, intra- and intermolecular radial distribution functions are shown in Figure 11. The intramolecular distances were determined using a least squares fitting procedure and resulted in 1.19±0.01 Å, 1.47±0.03 Å, 2.12±0.02 Å and 2.26±0.02 Å for the N–O, C–N, O–O and C–O bonds, respectively. Least-squares refined values for the X–H distances were: 1.07±0.01 Å for C–H, 2.06±0.03 Å for N···H and 1.74±0.01 Å for H···H. The results are shown in Table 2. Previous neutron diffraction studies on various molecular liquids resulted in the following values: C–H distance in methyl group 1.06-1.09 Å[51], H···H distance in methyl group 1.73-1.79 Å[51d,52] and N···H distances in formamide 2.04-2.09 Å.[53]

Overall, the intramolecular distances obtained by least squares fitting, agree fairly well with the similar distances in other compounds determined from neutron diffraction data. They agree also with the results of our X-ray diffraction study on proteated nitromethane. The intermolecular distribution function determined from the neutron data is however rather featureless, and just like in the case of the X-ray diffraction, no suggestions could have been made on the intermolecular structure of the nitromethane on the basis of neutron diffraction data. In *Section 5* a comparison of the total radial intermolecular distribution function obtained from molecular dynamics simulation and neutron diffraction is described.

*5. Comparison of experiments and simulations*

The radial distribution functions obtained from the X-ray and neutron diffraction experiments are compared with those obtained from simulation in Figures 9 and 11, respectively.

The total structure function relevant to the liquid structure (not including the intramolecular contribution) has been calculated from the partial radial distribution functions according to the equation

$$H(k) = \sum_{\alpha \geq \beta} \sum \frac{(2-\delta_{\alpha\beta})x_\alpha x_\beta f_\alpha f_\beta h_{\alpha\beta}(k)}{M(k)} \qquad (6)$$

where $f_\alpha$ is the scattering length or scattering factor of the α-type atom (which depend on $k$ in the case of X-ray diffraction), and $x_\alpha$ is the mole fraction of the α atom. $h_{\alpha\beta}(k)$ is defined according to the following equation:

$$h_{\alpha\beta}(k) = 4\pi\rho \int_0^{r_{max}} r^2 (g_{\alpha\beta}(r)-1) \frac{\sin(kr)}{kr} dr \qquad (7)$$

The total radial distribution function is defined as the Fourier transform of the structure function.

Figure 12a shows the contributions of each interaction to the total intermolecular radial distribution function, obtained from molecular dynamics simulation. It can be observed that the three interactions, C–X, O1–X and N–X contribute in nearly the same ratio to the total intermolecular radial distribution function. That's the reason for the noteworthy uncertainty of the X-ray diffraction method in determination of intermolecular structure of liquid nitromethane and the only suggestion could be the comparison between the X-ray diffraction and the theoretical radial distribution functions.

The agreement between the radial distribution function obtained by X-ray diffraction and the theoretical radial distribution function is very good, meaning that the average picture of the structure liquid nitromethane obtained by molecular dynamics simulation is confirmed by X-ray diffraction.

Figure 12b shows the contributions of each interaction to the total intermolecular radial distribution function, obtained from molecular dynamics simulation, for the case of neutron data. It can be observed that the situation is even worse than in the case of X-ray diffraction because all the interactions considered, except O-O, contribute with about 20% to the total intermolecular radial distribution function and there is not a real chance to resolve them. Comparing the total radial distribution functions resulted from neutron diffraction experiment and the simulation, a slight shift between them can be observed, which may originate from the not perfect separation of the inter- and intramolecular parts of the experimental structure factor.

## 6. Conclusions

A combined theoretical and experimental study of liquid nitromethane was carried out. The results of molecular dynamics simulations and diffraction experiments were compared in order to obtain a more reliable picture of the structure of the liquid. Besides the limitations of the applied methods are discussed.

The diffraction methods are known for the solution chemists as "direct methods for structural determination". This would mean that the parameters obtainable from radial distributions are characteristic of local structures in the liquid directly and no further speculation is needed for extracting them from the experimental data. However, the nitromethane is a very good example when one can see that the local structure of the liquid cannot be determined only on the basis of the diffraction data.

Quantum chemical calculations showed that there are two different conformers *eclipsed* and *staggered* of nitromethane molecule, with nearly the same total energy. The rotation barrier between these two minima is less then 0.01 kcal/mol, therefore the rotation of $NO_2$ group around the C–N bond is almost free. The optimized geometry for the dimer of nitromethane was obtained to be molecules shifted anti-parallel and it has been found that in the nitromethane dimer the CH···O interaction cannot be considered as H-bonded interaction.

The essential features of the RDFs in molecular dynamics and Car Parrinello simulations agree with each other and are similar to the results of earlier classical computer simulations. Car-Parrinello method revealed us that the molecular structure in the liquid state does not change significantly compared to the gas phase. The angle distribution of the HCNO dihedral angle, as obtained from the Car-Parinello simulation, show a uniform distribution between 0 and $\pi/6$, indicating that in the liquid state the rotation of $CH_3$ group is free. According to the molecular dynamics simulation study no predominated occurrence of nitromethane dimers in the liquid nitromethane could be detected, however the position and orientation of the nearest neighbor molecules resembles to some extent the configuration of the theoretically calculated energy-minimized dimer. Neighboring C–N bond vectors slightly prefer an anti-parallel orientation over a parallel one and there is no evidence for parallel or head-to-tail configurations in the liquid. In the liquid state CH···O type weak hydrogen bonds could not be detected.

The radial distribution function of MD simulation and X-ray diffraction agree very well. The small discrepancy between the radial distribution function obtained by simulation and neutron diffraction results may be due to both the potential model applied in the simulation and to the experimental uncertainties but the theoretical and experimental findings are in general accordance.

It is important to emphasize that while X-ray diffraction and neutron diffraction fails in determination of bulk structure, both methods are performing very well in determination of intramolecular distances. It should be emphasized, that we have obtained structural information of liquid nitromethane on the detailed analysis of the molecular dynamics simulations and comparison with diffraction experiments, therefore for deeper understanding of the liquid structure it is advisable to apply simultaneous theoretical and experimental methods.

**Acknowledgement**

The research was supported by projects NAP VENEUS05 OMFB-00650/2005 and the Hungarian Scientific Research Funds (OTKA), project numbers F 67929, K 68498. Support from the Hungarian/Austrian WTZ collaboration project A9/2004 is gratefully acknowledged. The authors wish to thank to Marie-Claire Bellisent-Funel her help with the neutron diffraction measurements on the diffractometer 7C2 of the Laboratoire Léon Brillouin CEA-SACLAY.

**Figure Captions**

Figure 1. Ball and stick representation of nitromethane conformers. a: eclipsed, b: staggered.

Figure 2. Ball and stick representation of dimer structure.

Figure 3. Partial radial distribution functions obtained from simulation. Molecular dynamics simulation: solid line, Car-Parrinello: dashed line.

Figure 4. Angle distribution of the HCNO dihedral angle, obtained in Car-Parinello simulation.

Figure 5. Distance dependent angular distribution function. Θ: angle between the dipole moment vector of center and neighboring molecules.

Figure 6. Orientational correlation functions: nitromethane (solid line) compared to acetonitrile (dashed line).

Figure 7. Distance dependent angular distribution function. Φ: angle between the dipole moment of central molecule and the center-center vector. a: acetonitrile, b: nitromethane.

Figure 8. Structure functions *h(k)* multiplied by *k* for nitromethane obtained by X-ray diffraction. Circles: experimental values, solid line: fitted values.

Figure 9. Radial distribution functions for nitromethane obtained from X-ray diffraction. Circles: experimental values; dashed line: intramolecular contribution; open circles: intermolecular contribution; solid line: intermolecular contribution obtained from MD simulation.

Figure 10. Total differential cross section and structure function (inset) of nitromethane obtained from neutron diffraction experiment (cross and open circles: experimental values, solid line: fit by the MCGR procedure).

Figure 11. Radial distribution function g(r) obtained from the neutron diffraction. Circles: experimental values; dashed line: intramolecular contribution; open circles: intermolecular contribution; solid line: intermolecular contribution obtained from MD simulation.

Figure 12. Contribution of different interactions to the total radial distribution function measured by a) X-ray diffraction and b) neutron diffraction.

Table 1.

Characteristic values for eclipsed and staggered conformers of nitromethane molecule obtained with MP2 and B3LYP methods compared to experimental data from literature. Atom-atom distances are given in Å, frequencies in cm$^{-1}$. (Subscripts *s* means symmetric, *as* means antisymmetric and *def* means deformation)

|  | exp. | eclipsed | | staggered | |
|---|---|---|---|---|---|
|  |  | MP2 | B3LYP | MP2 | B3LYP |
| $r_{C-N}$ | 1.489[a] | 1.497 | 1.503 | 1.491 | 1.503 |
| $r_{N-O}$ | 1.224[a] | 1.230 | 1.2209 | 1.230 | 1.221 |
| $r_{C-H1}$ | 1.089[a] | 1.087 | 1.086 | 1.0868 | 1.0886 |
| $r_{C-H2}$ | 1.089[a] | 1.090 | 1.090 | 1.091 | 1.0853 |
| $r_{C-H3}$ | 1.089[a] | 1.087 | 1.086 | 1.0868 | 1.0886 |
| $\angle(O_1NO_2)$ | 125.3[a] | 125.6 | 125.6 | 125.8 | 125.6 |
| $\angle(HCNO_1)$ |  | -179.9 | -179.1 | 26.9 | 27.0 |
| $\nu_{as}$ (CH$_3$) | 3048[b] | 3254 | 3197 | 3254 | 3198 |
| $\nu_{as}$ (CH$_3$) | 3048[b] | 3230 | 3166 | 3230 | 3165 |
| $\nu_s$ (CH$_3$) | 2965[b] | 3120 | 3075 | 3120 | 3076 |
| $\nu_{as}$ (NO$_2$) | 1582[b] | 1778 | 1623 | 1778 | 1624 |
| $\nu_s$ (NO$_2$) | 1413[b] | 1441 | 1428 | 1439 | 1428 |
| $\nu_{s,def}$ (CH$_3$) | 1488[b] | 1500 | 1474 | 1498 | 1475 |
| $\nu_{as,def}$ (CH$_3$) | 1449[b] | 1491 | 1464 | 1494 | 1465 |
| $\nu_{s,def}$ (CH$_3$) | 1384[b] | 1419 | 1400 | 1419 | 1399 |

[a] ref. 37, [b] ref. 38.

Table 2.

Structural and energetic parameters for nitromethane dimer obtained with MP2 and B3LYP methods. Atom-atom distances are given in Å, frequencies in cm$^{-1}$. (Subscripts *s* means symmetric, *as* means antisymmetric and *def* means deformation)

|  | MP2 | B3LYP |
|---|---|---|
| $-\Delta E$ | 7.34 | 4.45 |
| $-\Delta E_{BSSE}$ | 4.73 | 4.16 |
| $r_{C-N}$ | 1.493 | 1.503 |
| $r_{N-O}$ | 1.230 | 1.2209 |
| $r_{C-H1}$ | 1.087 | 1.086 |
| $r_{C-H2}$ | 1.093 | 1.090 |
| $r_{C-H3}$ | 1.087 | 1.086 |
| $\angle(O_1NO_2)$ | 125.9 | 125.7 |
| $\angle(HCNO_1)$ | -89.2 | -85.8 |
| $r_{CH\cdots O}$ | 2.45 | 2.43 |
| $\nu_{as}(CH_3)$ | 3253 | 3198 |
| $\nu_{as}(CH_3)$ | 3231 | 3164 |
| $\nu_s(CH_3)$ | 3119 | 3072 |
| $\nu_{as}(NO_2)$ | 1770 | 1617 |
| $\nu_s(NO_2)$ | 1443 | 1432 |
| $\nu_{s,def}(CH_3)$ | 1499 | 1473 |
| $\nu_{as,def}(CH_3)$ | 1496 | 1469 |
| $\nu_{s,def}(CH_3)$ | 1429 | 1403 |

Table 3.

Characteristic values for the radial distribution functions $g_{\alpha\beta}(r)$. $n_{\alpha\beta}$ is the running integration number. Atom-atom distances are given in Å.

| bond type | $r_{max}$ | $g_{\alpha\beta}(r_{max})$ | $r_{min}$ | $n_{\alpha\beta}(r_{min})$ |
|---|---|---|---|---|
| C-C | 5.17 | 1.89 | 6.57 | 13.2 |
| C-O | 3.42 | 1.82 | 4.57 | 3.7 |
| C-N | 4.27 | 2.04 | 5.07 | 5.8 |
| C-H | 4.77 | 1.35 | 5.42 | 6.8 |
| O-O | 3.47 | 1.06 | 4.37 | 2.5 |
| O-N | 4.47 | 1.29 | 5.47 | 3.4 |
| O-H | 2.92 | 1.15 | 3.67 | 4.7 |
| N-N | 5.12 | 1.52 | 6.52 | 12.4 |
| N-H | 3.62 | 1.27 | 4.37 | 8.7 |

Table 4.

Structural parameters from the X-ray and neutron diffraction refinement with estimated errors in the last digits. *n* is the coordination number. Distances (*r*) and their mean-square deviations (σ) are given in Å.

| Bond type | r | σ | n |
|---|---|---|---|
| *X-ray diffraction* | | | |
| C-N | 1.49(1) | 0.08(2) | 1 |
| C-O | 2.32(1) | 0.10(1) | 2 |
| N-O | 1.22(1) | 0.05(1) | 2 |
| O-O | 2.17(2) | 0.10(2) | 1 |
| *Neutron diffraction* | | | |
| C-N | 1.47(2) | 0.12(1) | 1 |
| C-O | 2.26(2) | 0.13(2) | 2 |
| N-O | 1.19(1) | 0.09(2) | 2 |
| O-O | 2.12(2) | 0.12(1) | 1 |
| C-H | 1.07(1) | 0.07(1) | 3 |
| N-H | 2.06(3) | 0.13(2) | 3 |
| H-H | 1.74(1) | 0.13(1) | 2 |

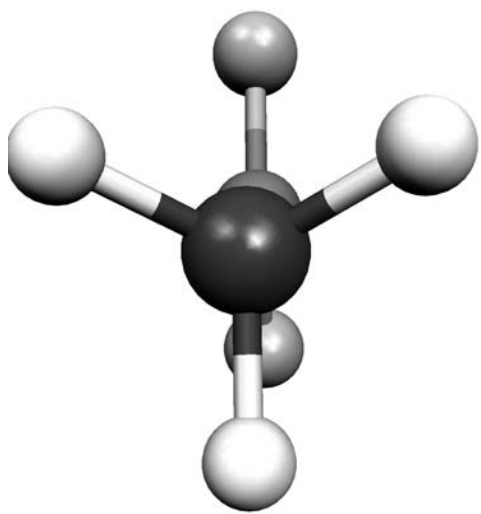     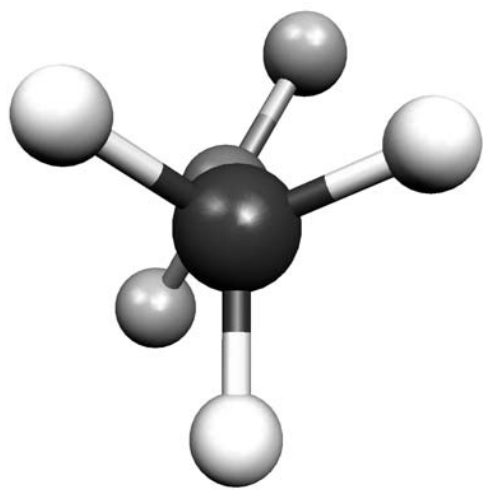

a                                  b

Figure 1.

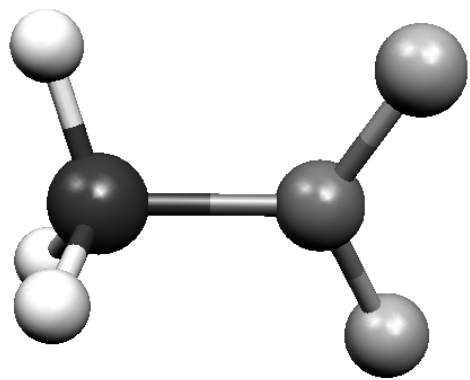

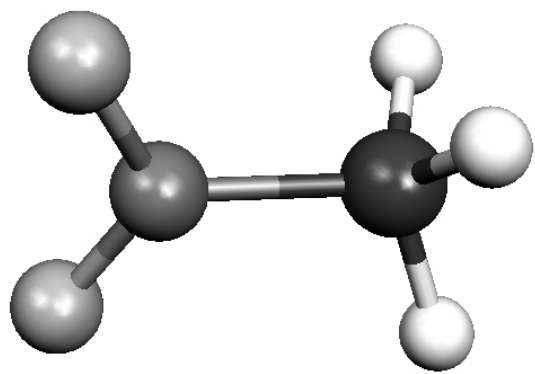

Figure 2.

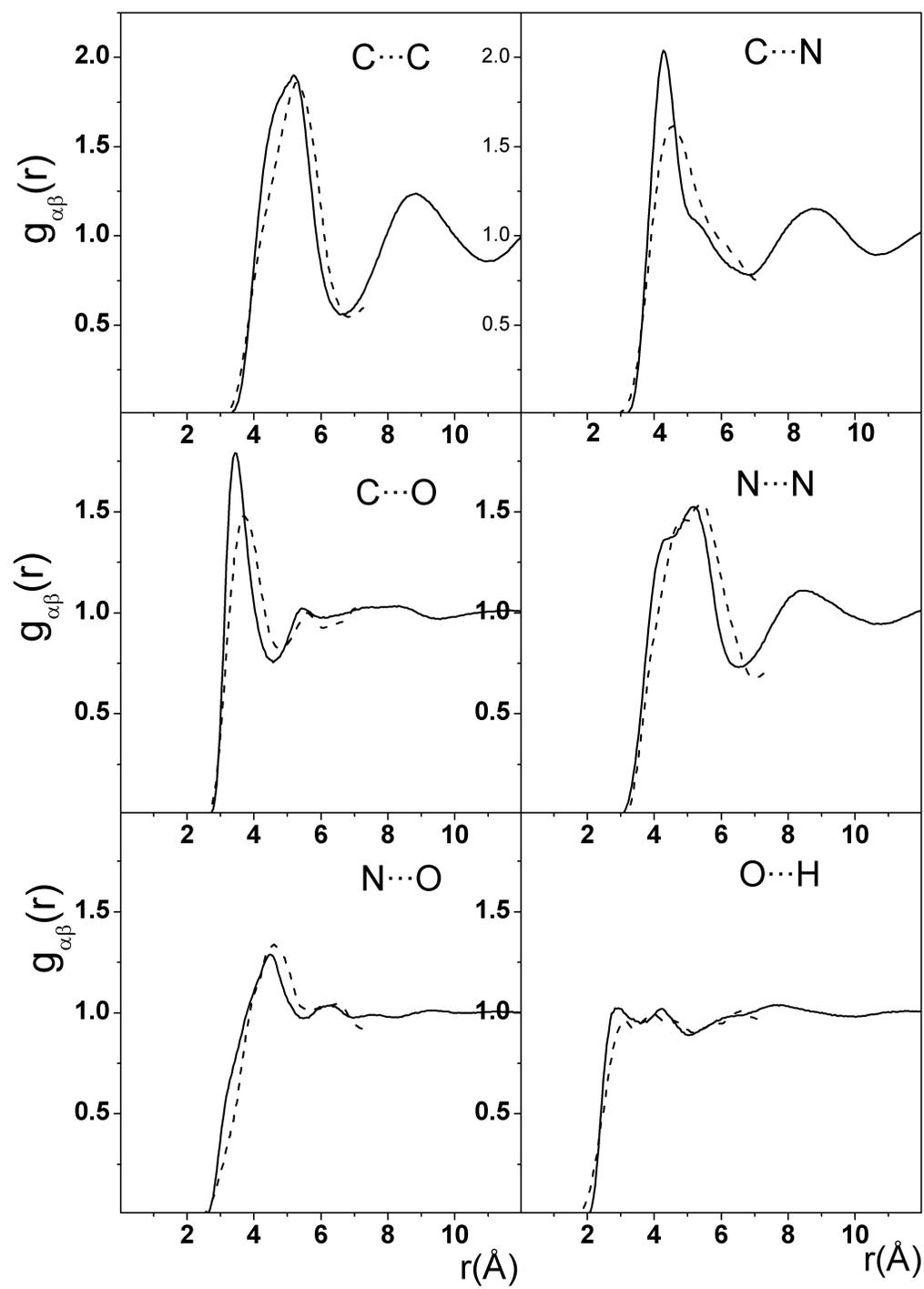

Figure 3.

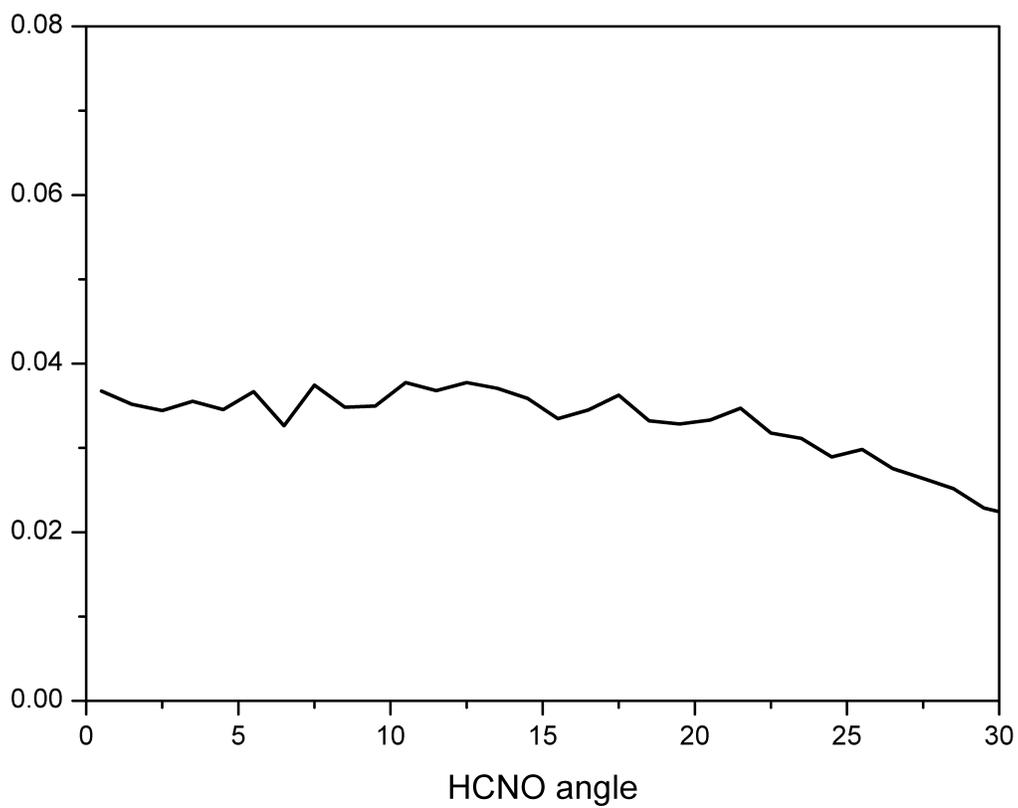

Figure 4.

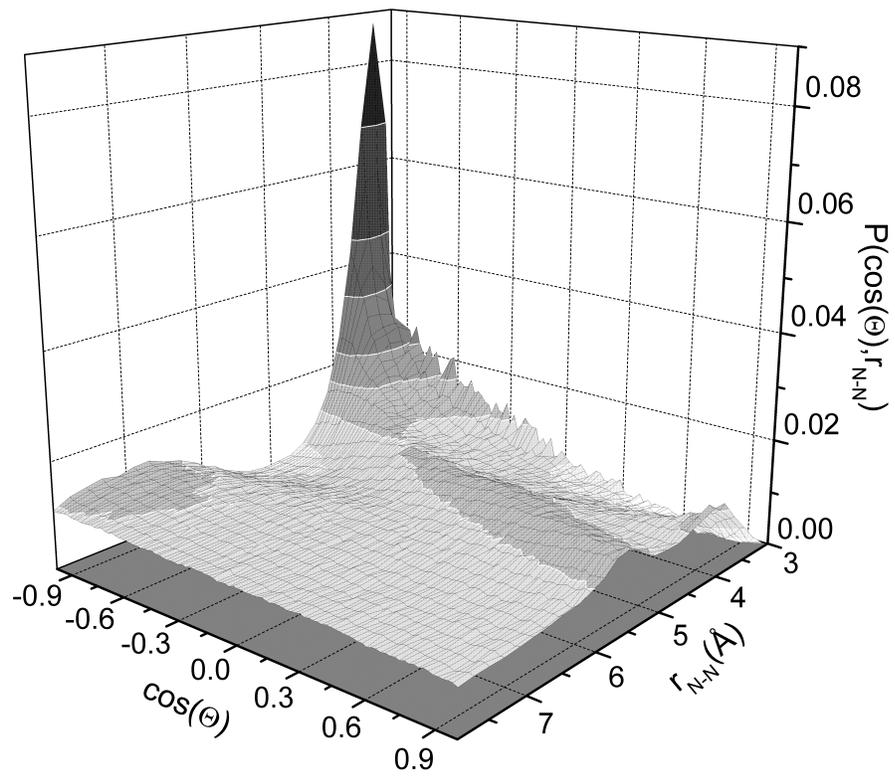

Figure 5.

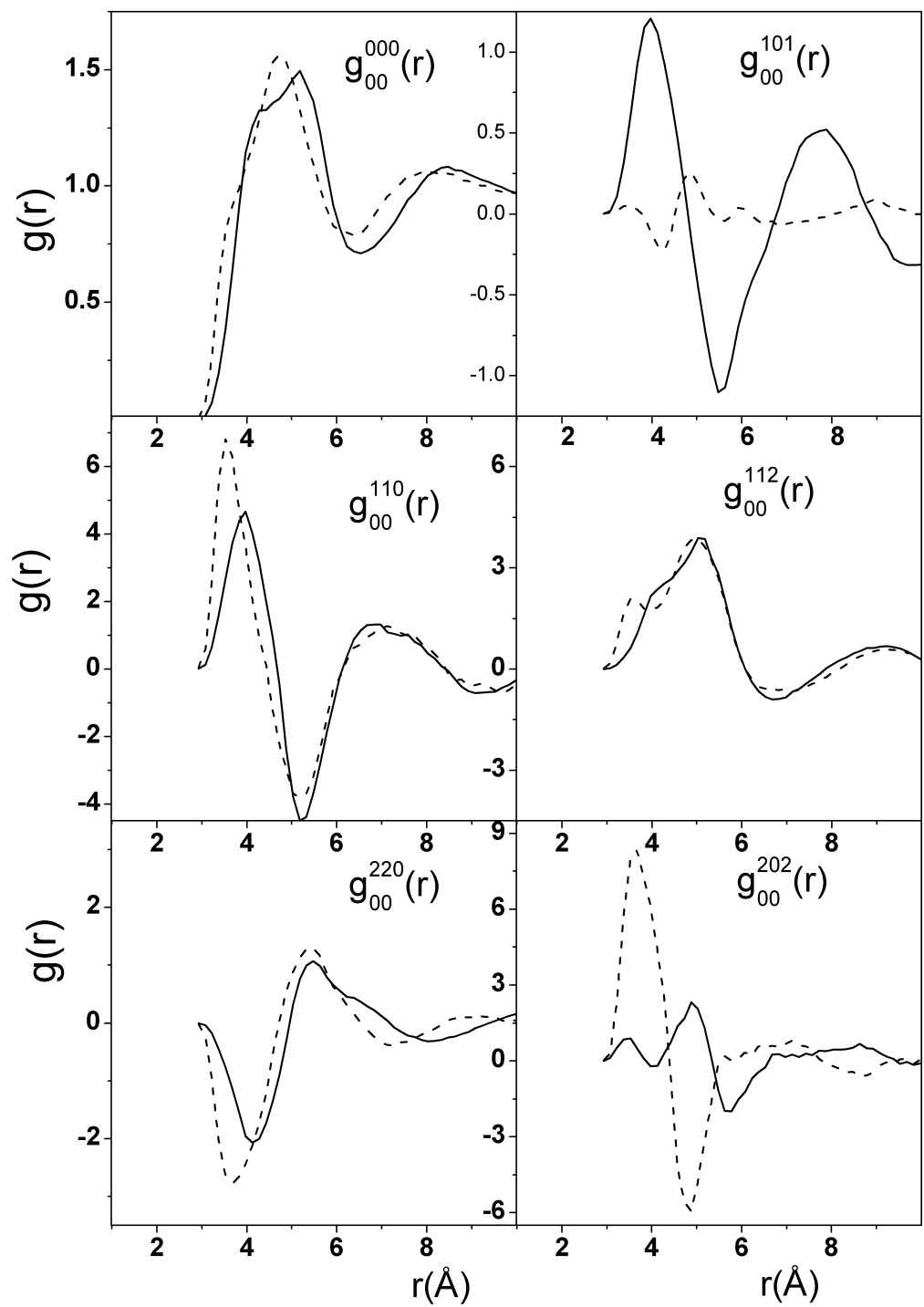

Figure 6.

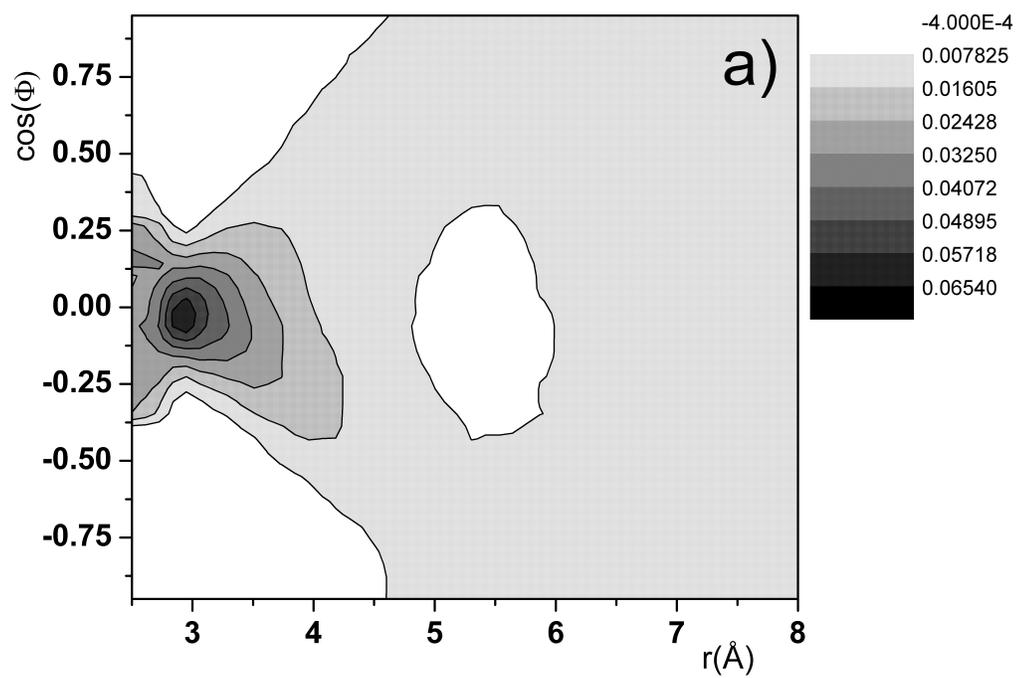

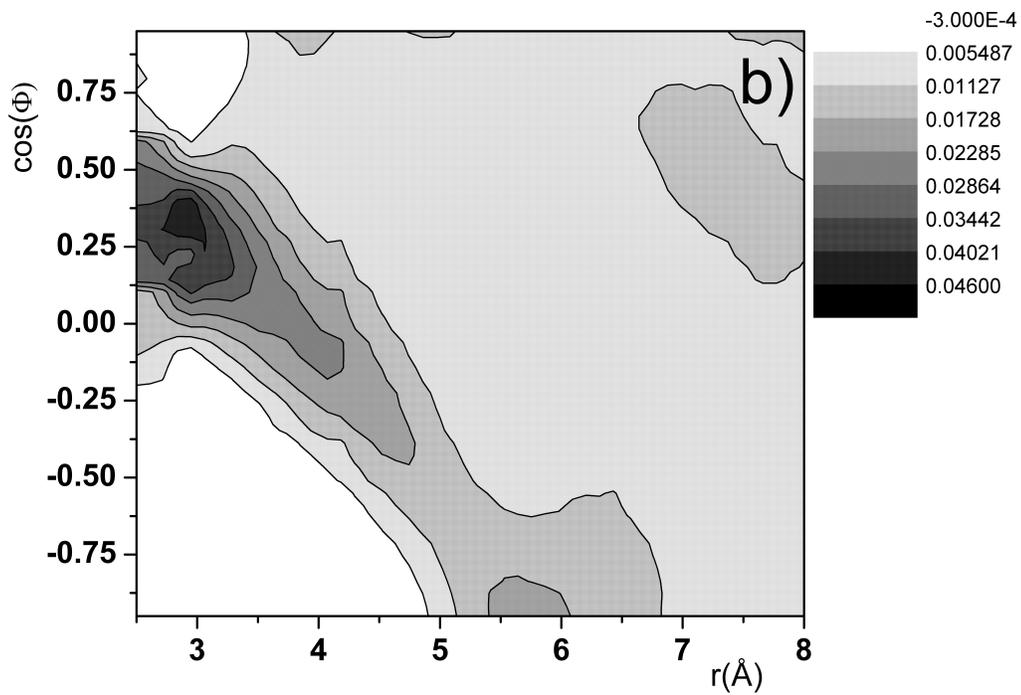

Figure 7.

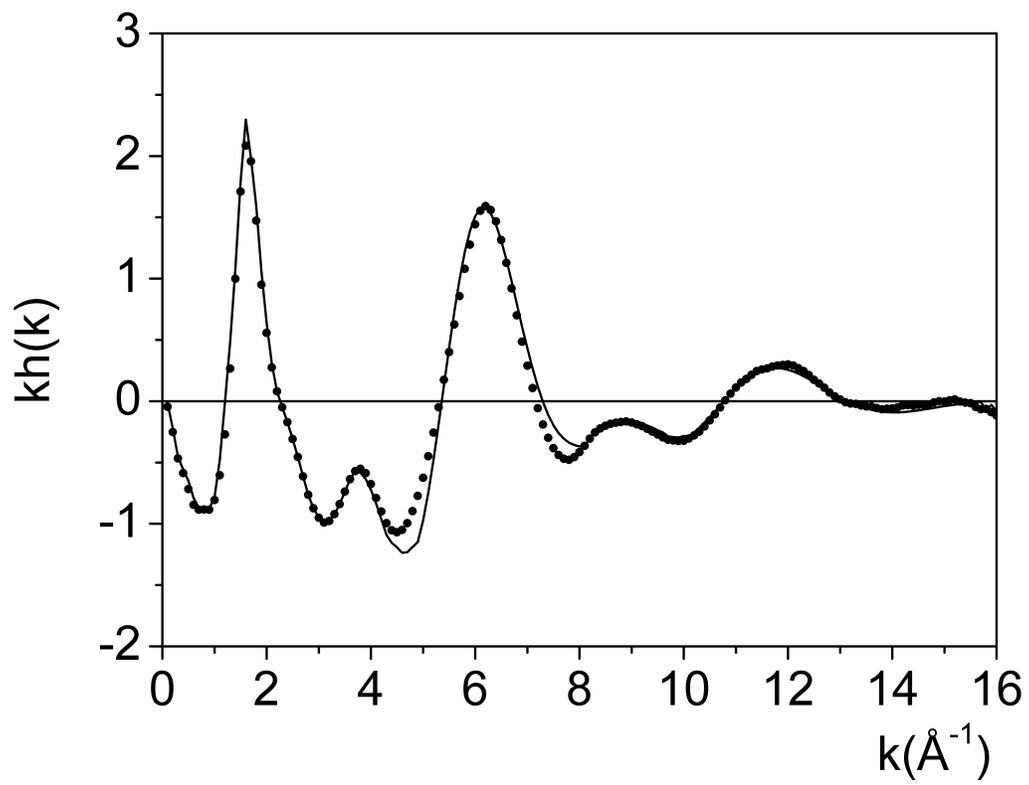

Figure 8.

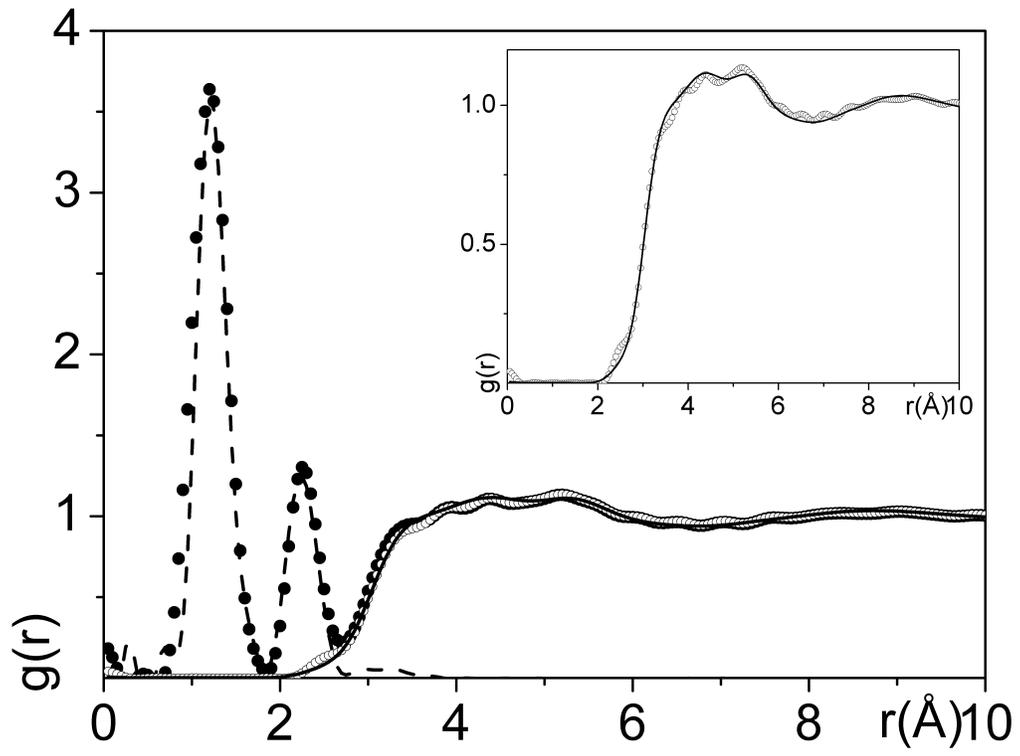

Figure 9.

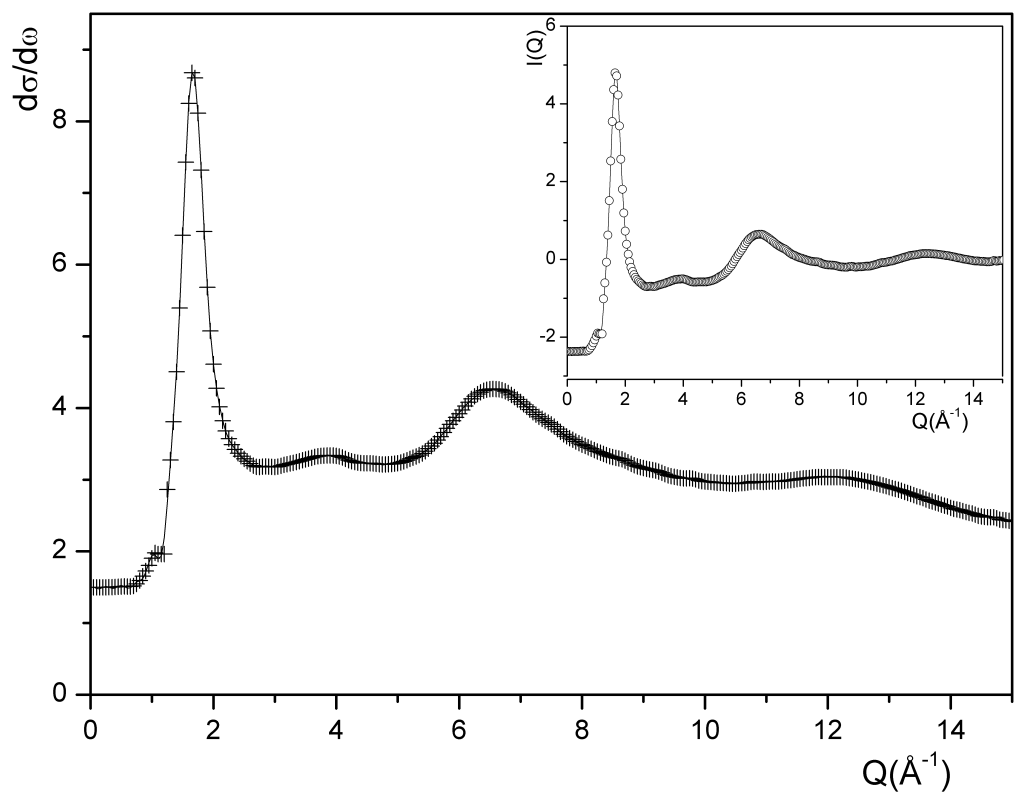

Figure 10.

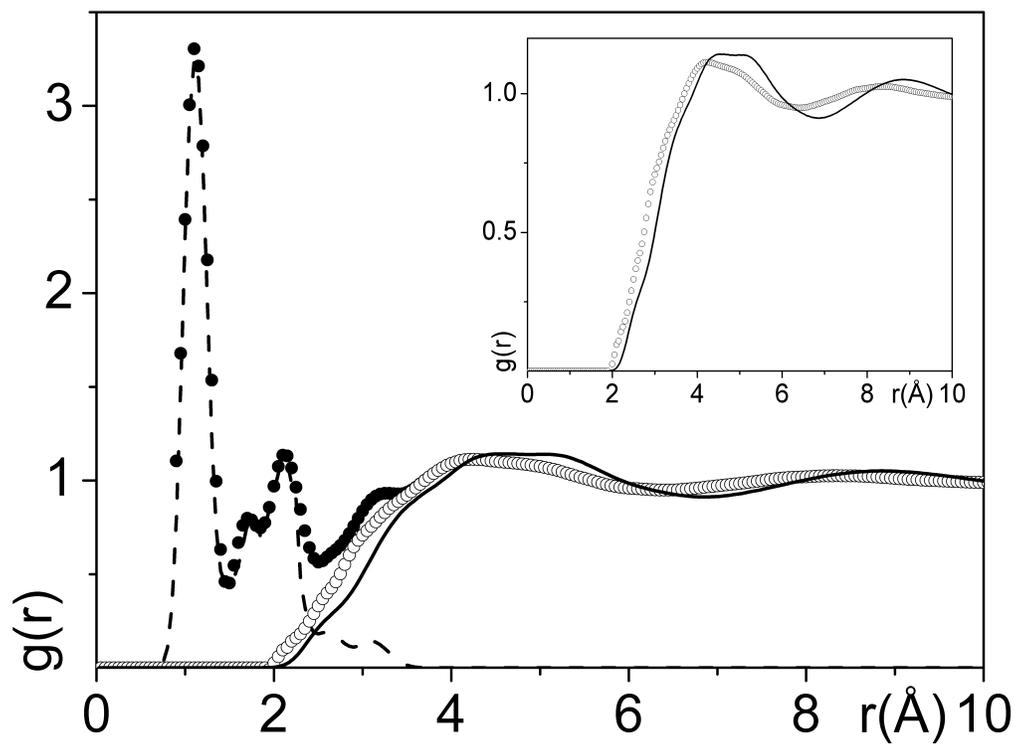

Figure 11.

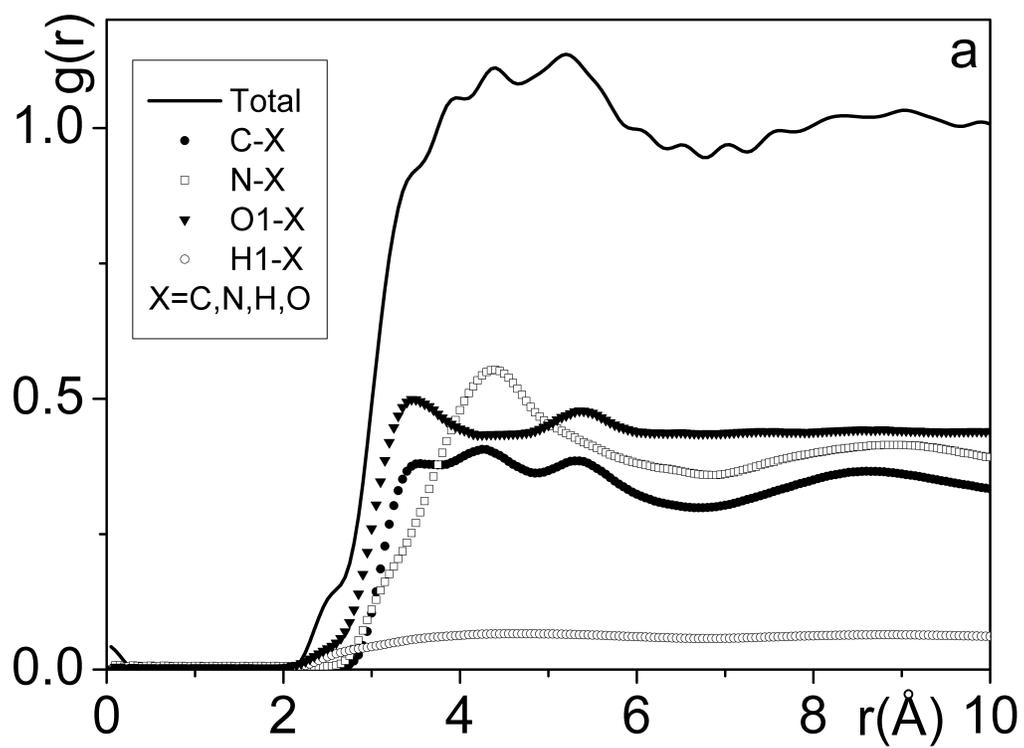

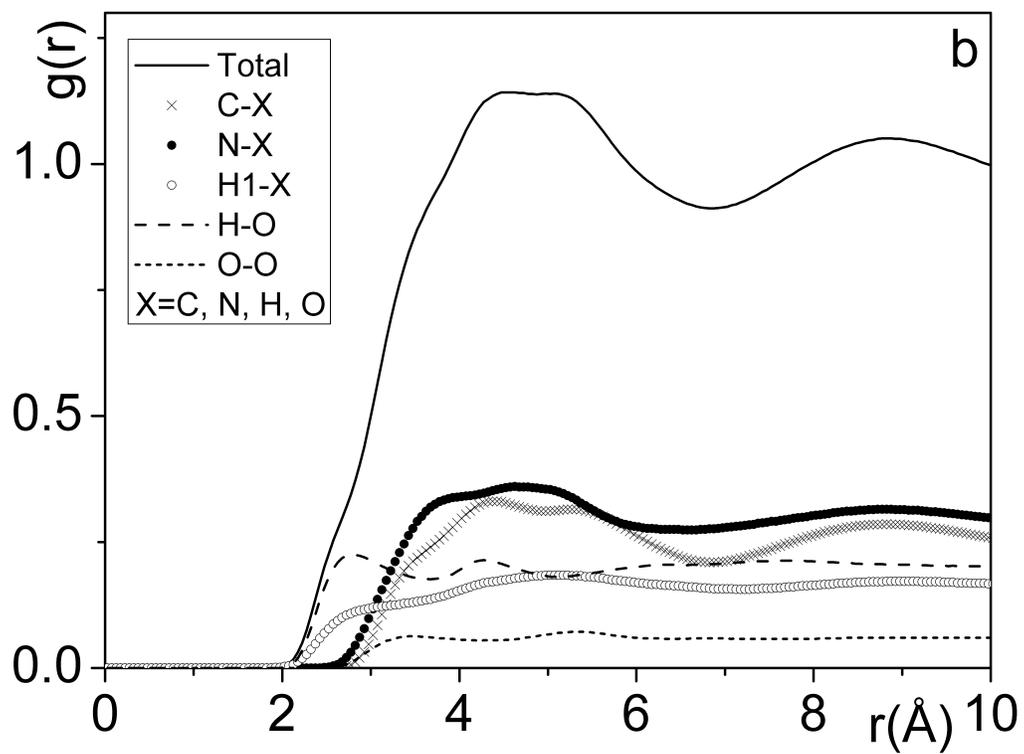

Figure 12.